\documentstyle[12pt]{article}

\input{psfig}

\font\mybb=msbm10 at 11pt
\def\bb#1{\hbox{\mybb#1}}

\newlength{\extraspace}
\newlength{\extraspaces}
\newcommand{\bq}{\begin{eqnarray}
\addtolength{\abovedisplayskip}{\extraspaces}
\addtolength{\belowdisplayskip}{\extraspaces}
\addtolength{\abovedisplayshortskip}{\extraspace}
\addtolength{\belowdisplayshortskip}{\extraspace}}
\newcommand{\eq}{\end{eqnarray}}

\newcommand{\newsection}[1]
{\vspace{5mm}
\pagebreak[3]
\addtocounter{section}{1}
\setcounter{equation}{0}
\setcounter{subsection}{0}
\setcounter{footnote}{0}
\begin{flushleft}
{\large\bf \thesection. #1}
\end{flushleft}
\nopagebreak
\medskip
\nopagebreak}

\begin{document}

\addtolength{\baselineskip}{.8mm}

\thispagestyle{empty}

\begin{flushright}
{\sc OUTP}-96-06P\\
 Revised, June 1996\\
 hep-th/9602062
\end{flushright}
\vspace{.3cm}

\begin{center}
{\large\sc{BOUNDARY CONDITIONS AND HETEROTIC
CONSTRUCTION IN TOPOLOGICAL MEMBRANE THEORY} }\\[15mm]

{\sc  Leith Cooper\footnote{leith@thphys.ox.ac.uk} 
and Ian  I. Kogan\footnote{kogan@thphys.ox.ac.uk}} \\[2mm]
{\it Theoretical Physics, 1 Keble Road,
       Oxford, OX1 3NP, UK} \\[15mm]

{\sc Abstract}

\begin{center}
\begin{minipage}{14cm}

Using the topological membrane approach to string theory, we
suggest a geometric origin for the heterotic string. We show
how different membrane boundary conditions lead to different
string theories. We discuss the construction of closed oriented 
strings and superstrings, and demonstrate how the heterotic
construction naturally arises from a specific choice of boundary
conditions on the left and right boundaries of a cylindrical 
topological membrane.

\end{minipage}
\end{center}

\end{center}

\noindent

\vfill
\newpage
\pagestyle{plain}
\setcounter{page}{1}
\stepcounter{subsection}
\newsection{Introduction}
\renewcommand{\footnotesize}{\small}

 The  heterotic  string construction \cite{het,freund},
 in which left and right sectors are taken from  different string 
theories (critical bosonic and superstrings in the original papers
\cite{het}), has no clear  geometrical meaning within string theory
itself.  Although the left
and right sectors may each separately be viewed geometrically as a 
two-dimensional conformal field theory  on a two-dimensional random 
surface (the worldsheet), the heterotic construction remains 
somewhat hybrid. The purpose of this letter is to suggest a 
geometric origin for the heterotic string using the topological 
membrane ({\sc tm})  approach to string theory 
\cite{membrane1,membrane2}. It has recently been shown that 
dualities exist between different types of string theories,
including dualities betwen heterotic and non-heterotic strings
\cite{duality}. These dualities relate the strong coupling behaviour
of one string theory to another at weak coupling. Hence, a weakly
coupled heterotic string is dual to some strongly coupled string
theory. From this point of view, we hope that in obtaining a better
understanding of the geomtric nature of the heterotic string at weak 
coupling, we might gain further insight into the strong coupling
behaviour of other string theories.

The basic idea behind the {\sc tm} approach to string theory 
 is to fill in the string worldsheet and 
view it as the boundary of a three-manifold. If inside the 
three-manifold there 
is a  topologically massive gauge theory ({\sc tmgt}) \cite{tmgt},
it induces the chiral gauged Wess-Zumino-Novikov-Witten ({\sc wznw})
action on the boundary worldsheet. This picture is based upon the
remarkable connection between Chern-Simons and conformal field
theories discovered in  \cite{witten} and elaborated in
\cite{ms}. Based partly upon the earlier work of \cite{km}, it was 
then found \cite{membrane1,membrane2} that the boundary degrees  
of freedom are also induced  in {\sc tmgt}, where massive vector 
particles propagate in the bulk.
It is interesting to note that the same  picture  can be used to describe
edge excitations in quantum Hall systems with boundaries. A nice
theory of chiral edge states based on this approach was 
constructed in \cite{edge}.  In some sense, gapless edge states
in  planar condensed matter  systems are the  ``solid-state model'' 
for chiral string sectors.

The basic property allowing a three-dimensional string construction 
is the {\em chiral} nature of the induced conformal fields on the 
boundary. To obtain a realistic closed string theory, of course, 
we need both left- and right-movers. We can accomplish this by using
an annular or cylindrical topological membrane with left ({\sc l}) and
right ({\sc r}) boundaries. The bulk {\sc tmgt} then induces a chiral
{\sc wznz} model with left-movers on {\sc l} and right-movers on
{\sc r}. The crucial idea is that we can independently excite the
left and right sectors by choosing different boundary conditions 
on the left and right boundaries --- and this is possible only in 
{\sc tmgt}, where the gauge field propagates in the bulk, and not in 
a pure Chern-Simons theory where there are no such propagating degrees
of freedom in the bulk, only topological ones. This is why the
topological membrane is so important: by specifying different membrane
boundary conditions (which we shall discuss in detail later), we can
construct different closed strings, including both type II theories
and also the heterotic string.

We begin by showing how {\sc tmgt} induces chiral bosonic
edge states and how their appearance depends upon the boundary
conditions for the gauge field in the bulk. In Section\,3
we show how the supersymmetric topological membrane leads to the
appearance of fermions on the boundary. In this way we incorporate
worldsheet {\sc susy}. In Section\,4 we show how 
different membrane boundary conditions lead to different string 
constructions. 
For each of these constructions we calculate the spectrum in the
bulk of the membrane and the resultant Casimir energy. As we shall
see in Section\,5, this energy distinguishes between
the heterotic construction and all the others. We conclude by
comparing our results with recent work in $M$-theory.
For simplicity, in this letter we shall only consider the abelian 
theory. We hope to give the non-abelian generalization in a future
publication.

\newsection{Bosonic edge states from TMGT}  

Let us consider how {\sc tmgt} induces chiral bosonic edge 
states and how their appearance depends upon the boundary
conditions for the gauge field. There are numerous papers
 in which this question was discussed, here we  follow the
 lines of \cite{carlip1991a}.
 The abelian theory  is defined
by the action
\begin{equation}
S_1=-\frac{1}{4\gamma}\int_{\cal M}\sqrt{-g}\,F_{\mu\nu}F^{\mu\nu}
  + \frac{k}{16\pi}\int_{\cal M}\epsilon^{\mu\nu\alpha}F_{\mu\nu}A_\alpha\, .
\label{action}
\end{equation}
If the three-manifold $\cal M$ is compact, then varying the action with
respect to $A_\nu$ yields the equations of motion 
\begin{equation}
\partial_\mu F^{\mu\nu} + \frac{m}{2}\epsilon^{\nu\alpha\beta}
F_{\alpha\beta}=0\,, 
\label{eqmotion2}
\end{equation}
where $m=\gamma k/4\pi$. The Bianchi identity $\partial_\mu F^\mu=0$ 
on the dual field strength $F^\mu=\mbox{$\frac{1}{2}$}\epsilon^
{\mu\alpha\beta}F_{\alpha\beta}$ follows immediately from 
(\ref{eqmotion2}). Moreover,
taking the curl of (\ref{eqmotion2}) shows that the gauge field is
indeed massive:
\begin{equation}
(\partial^2-m^2)F^\mu=0\, ,
\label{wave2}
\end{equation}
with one propagating degree of freedom (recall that a gauge field
in $d$-dimensional space-time descibes $d-2$ degrees of freedom, i.e. 
$ 3-2 =1 $ in our case).

Now consider the case when ${\cal M}$ has a boundary. We assume the
boundary $\partial{\cal M}$ contains the timelike direction and use 
light-cone coordinates $x^\pm=x^0\pm x^1$ for the directions on the 
boundary and $x^2=x^\perp$ for the perpendicular direction. Under
arbitrary variations, $S_1$ has no extrema:
\begin{equation}
\delta S_1=\int_{\cal M}\left(\frac{1}{\gamma}\partial_\mu 
 F^{\mu\nu}-\frac{k}{8\pi}\epsilon^{\mu\nu\lambda}F_{\mu\lambda}
 \right)\delta A_\nu+\int_{\partial\cal M}n_\perp\Pi^{\perp i}
 \delta A_{i} 
\label{vari}
\end{equation}
where $n_\perp$ is a unit-vector normal to the boundary, $i=+,-$ and
\begin{equation}
\Pi^{\perp i}=-\frac{1}{\gamma} F^{\perp i} + \frac{k}{8\pi}
 \epsilon^{\perp ij}A_j\,.
\label{mom}
\end{equation}
To obtain a sensible path integral, with a well-defined classical
limit, the boundary terms must somehow vanish. This problem is 
equivalent to choosing the necessary boundary conditions in order
to solve the bulk equations of motion. 
\begin{itemize}
\item {\bf C boundary conditions:}
\end{itemize}
We may choose to fix $A_+$
(upto gauge transformations) on $\partial\cal M$, while allowing
$A_{-}$ to vary. To eliminate the surface term in (\ref{vari}) we
must therefore add 
\begin{equation}
S_2=-\int_{\partial\cal M}\Pi^{\perp-}A_{-}
\end{equation}
and fix $\delta\Pi^{\perp-}=0$ on $\partial{\cal M}$. We denote these
boundary conditions by {\sc C=Conformal} and will now show that they
lead to the appearance of conformal degrees of freedom on
$\partial{\cal M}$.

Although the combined action $S=S_1+S_2$ can now be extremized, it is
no longer gauge invariant. The point is, however, that with the
addition of $S_2$ and upon gauge fixing, some gauge degrees of freedom
become dynamical on $\partial\cal M$. This is best seen via the 
Faddeev-Popov gauge-fixing procedure. Explicitly, under the $U(1)$ 
gauge transformation
$A_\mu=\bar A_\mu+\partial_\mu\theta$, we have
\begin{equation}
\delta S[\theta,\bar A_+,\bar\Pi^{\perp-}]=\frac{k}{8\pi}\int_{
\partial\cal M}\partial_+\theta\,\partial_{-}\theta +
\partial_{-}\theta\left(\bar A_+ -\mbox{$\frac{8\pi}{k}$}
\bar\Pi^{\perp-}\right)\, .
\label{wzw}
\end{equation}
In order to fix the gauge, we write the partition 
function as
\begin{equation}
{\cal Z}=\int{\cal D}A_\mu\left(\Delta_{\tiny{FP}}\int{\cal
 D}\theta\,\delta\Big(F(\bar A_\mu)\Big)\right)e^{iS[A_\mu]}
\end{equation}
where $F(\bar A_\mu)=0$ is the gauge-fixing condition and 
$\Delta_{\tiny{FP}}$ is the Faddeev-Popov determinant. 
Under the change of variables $A_\mu\rightarrow \bar A_\mu$, the 
partition function factorizes as
\begin{equation}
{\cal Z}=\int{\cal D}\bar A_\mu\,\Delta_{\tiny{FP}}\,
 \delta\Big(F(\bar A_\mu)\Big)\,e^{iS[\bar A_\mu]}\int{\cal D}\theta\,
 e^{i\delta S[\theta,\bar A_{+},\bar\Pi^{\perp-}]}\,.
\label{fp}
\end{equation}
The first term is the standard gauge-fixed path integral for 
{\sc tmgt}, describing a massive photon propagating in the bulk. 
The second term is a surface contribution describing a chiral 
scalar field $\theta$ coupled to $\bar A_+$ and $\bar\Pi^{\perp-}$
which are fixed by the boundary conditions, and so the bulk and 
surface contributions completely decouple. Now that we have 
gauge-fixed we may, without loss of generality, fix $\bar A_+$
and $\bar\Pi^{\perp-}$ to zero on $\partial{\cal M}$. Thus we
obtain the path integral for chiral bosonic edge states in
{\sc tmgt}:
\begin{equation}
{\cal Z}=\int{\cal D}\bar A_\mu\,\Delta_{\tiny{FP}}\,
 \delta\Big(F(\bar A_\mu)\Big)\,e^{iS[\bar A_\mu]}\int{\cal D}\theta\,
 e^{iS_{\tiny B}[\theta]}\, ,
\end{equation}
where
\begin{equation}
S_{\tiny B} =\frac{k}{8\pi}\int_{\partial\cal M}\partial_+\theta\,
\partial_{-}\theta
\label{bosonic}
\end{equation}
may be recognized as the bosonic string action in light-cone
coordinates. 
\begin{itemize}
\item $\tilde{\mbox{\bf C}}$ {\bf boundary conditions:}
\end{itemize}
Alternatively, we could choose to fix $\bar A_{-}$ on $\partial
{\cal M}$ while allowing $\bar A_+$ to vary. In order to extremize 
$S_1$ we must add
\begin{equation}
\tilde{S_2}=-\int_{\partial\cal M}\Pi^{\perp+}A_+
\end{equation}
and fix $\delta\Pi^{\perp+}=0$ on $\partial{\cal M}$. We denote
these boundary conditions by $\tilde{\mbox{\sc c}}$.
Applying the {\sc fp} gauge-fixing procedure we obtain
\begin{equation}
\tilde{S}_{\tiny B} =-\frac{k}{8\pi}\int_{\partial\cal M}
\partial_+\theta\,\partial_{-}\theta\,.
\label{bosonic2}
\end{equation}
Since parity \cite{tmgt} reverses the sign of $k$ and interchanges
$A_+\leftrightarrow A_{-}$, we see that {\sc c} and 
$\tilde{\mbox{\sc c}}$ boundary conditions are related by a 
parity transformation. Hence we need only consider {\sc c}
boundary conditions and take $k$ to be positive.
\begin{itemize}
\item {\bf N boundary conditions:}
\end{itemize}
Now consider the case where both $\bar A_+$ and $\bar A_{-}$ are 
fixed on $\partial{\cal M}$. We denote these boundary conditions by
{\sc N=Non-conformal} because they do not induce conformal edge
states. Since both $\bar A_{+}$ and $\bar A_{-}$ are fixed on 
$\partial{\cal M}$, the surface variations in (\ref{vari}) vanish
and so $S_1$ has a well-defined classical extrema. If we now gauge
fix just $S_1$ using the {\sc fp} procedure, the integral over the 
gauge group becomes
\begin{equation}
\int{\cal D\theta}\,e^{i\,\delta S_1[\theta,\bar A_\pm]}
 =\int{\cal D\theta}\,e^{\,{\mbox{$\frac{ik}{8\pi}$}\int_{\partial{\cal M}}
 \theta\epsilon^{ij}\partial_i\bar A_j}}\, \sim 
\delta\left(F_{+-} \right)\Big|_{\partial M}\, .
\label{perfectconductor}
\end{equation}
If $\theta$ is non-compact,  we get a $\delta$-functional 
which tells us that the boundary conditions for $\bar A_+$ and 
$\bar A_{-}$  must be compatible with the constraint $F_{+-} = 0$ 
on $\partial{\cal M}$. (Note that if $\theta$ is compact we get the 
constraint mod\,$2\pi$). This constraint holds when the boundary 
$\partial{\cal M}$ is perfectly conducting, for which the tangential 
component of the field strength must vanish. We conclude that {\sc n}
boundary conditions have no dynamical degrees of freedom on 
$\partial{\cal M}$. It is important to note that we can only impose 
{\sc n} boundary conditions in the full {\sc tmgt}: in pure 
Chern-Simons theory $[\bar A_+,\bar A_{-}]=i\hbar$, and so we cannot 
simultaneously fix both $\bar A_+$ and $\bar A_{-}$ on $\partial
{\cal M}$. 
 
\newsection{Fermionic edge states from SUSY TMGT} 

We now show how the supersymmetric topological membrane leads
to chiral fermionic edge states and, hence, worldsheet {\sc susy}.
The idea is simple: under {\sc susy} transformations, the action
is invariant only upto total derivatives --- in order to maintain
{\sc susy} for manifolds with boundaries, we must have chiral 
fermions on the boundary. In \cite{sakaiET1990a} and 
\cite{membrane2} it was shown how to obtain super-{\sc wznw} 
models from pure Chern-Simons theories. Here we consider the 
supersymmetric extension of abelian {\sc tmgt} using the 
conventions given in  Appendix A of \cite{sakaiET1990a}. 

{\sc susy tmgt} is described by the bulk action $S=S_1+S_3$
where
\begin{equation}
S_3 =  \frac{1}{2\gamma}\int_{\cal M}\bar\lambda\gamma^\mu
\partial_\mu\lambda-\frac{k}{8\pi}\int_{\cal M}\bar\lambda\lambda\, .
\label{sf}
\end{equation}
Ignoring surface terms for the moment, $S$ is invariant under the
$N=1$ {\sc susy} transformations:
\begin{eqnarray}
\delta A_\mu &=& -\bar\eta\gamma_\mu\lambda+\bar\eta\partial_\mu\psi
\label{susya}\\
\delta\lambda &=& \epsilon^{\mu\nu\rho}\partial_\nu A_\rho
\gamma_\mu\eta\, , \label{susyl}
\end{eqnarray}
where $\eta$ is a global Grassman parameter. The second term in 
(\ref{susya}) is equivalent to a gauge transformation parameterized 
by $\theta(x)=\bar\eta\psi(x)$ and, as we shall see, is needed
to maintain {\sc susy} when ${\cal M}$ has a boundary.

If $\partial{\cal M}\neq 0$, then the fermionic action $S_3$ 
has no extrema:
\begin{equation}
\delta S_3 = \int_{\cal M}\delta\bar\lambda\left(
\frac{1}{\gamma}\gamma^\mu\partial_\mu\lambda-\frac{k}{4\pi}\lambda
\right)+\frac{i}{2\gamma}\int_{\partial\cal M}(\lambda_{+}\delta
\lambda_{-}+\lambda_{-}\delta\lambda_{+})
\label{varif}
\end{equation}
and so we must fix either $\lambda_{-}$ or $\lambda_{+}$ on the
boundary to ensure a well-defined classical limit. Since
$\{\lambda_{-},\lambda_{+}\}=i\hbar$, we cannot simultaneously fix 
both $\lambda_{-}$ and $\lambda_{+}$ on $\partial{\cal M}$.
As we shall soon see, {\sc susy} dictates whether $\lambda_{-}$
or $\lambda_+$ is fixed on $\partial{\cal M}$. Secondly, since two 
$N=1$ {\sc susy} transformations generate translations in all
three directions, translation in the direction perpendicular to
the boundaries cannot be avoided unless we explicitly break the
$N=1$ {\sc susy} to an $N=1/2$ chiral {\sc susy}. Hence
we must set either $\eta_{+}=0$ or $\eta_{-}=0$. First consider
the case $\eta_+=0$. 
\begin{itemize}
\item[i)] $\eta_+ =0:$
\end{itemize}
The restricted {\sc susy} 
transformations are:
\begin{eqnarray}
\delta\bar A_+ & = & \!\!-i\eta_+\lambda_+ = 0 \label{rsusy1}\\
\delta\bar A_- & = &  i\eta_-\lambda_-     \label{rsusy2}\\
\mbox{$\delta\lambda\,\,\,$}  & = & 2F_{+-}\pmatrix{\,\eta_{-} \cr 0}
 + F_{\perp+}\pmatrix{0 \cr \,\,\eta_{-}}\, . \label{rsusy3}
\end{eqnarray}
We will show that {\sc c} boundary conditions combined with {\sc susy}
leads to the supersymmetric {\sc wz} model on $\partial{\cal M}$, while
for {\sc n} boundary conditions there are no such dynamical edge
states.
\begin{itemize}
\item {\bf N  boundary conditions:}
\end{itemize}
Recall from Section\,2 that {\sc n} boundary conditions are
$\bar A_+$ and $\bar A_{-}$ fixed on $\partial{\cal M}$ along with
the constraint (\ref{perfectconductor}) that $F_{+-}=0$ on 
$\partial{\cal M}$. Since $F_{+-}=0$ on $\partial{\cal M}$, 
it follows from Eq.\,(\ref{rsusy3}) that $\delta\lambda_{-}=0$ on 
$\partial{\cal M}$. Moreover, since $\bar A_{-}$ is fixed on 
$\partial{\cal M}$, Eq.\,(\ref{rsusy2}) implies that we must fix 
$\lambda_{-}=0$ on $\partial{\cal M}$. Hence the surface variations
in (\ref{varif}) vanish and $S_3$ has a well-defined classical
limit. Under the restricted {\sc susy} transformations (\ref{rsusy1})
-- (\ref{rsusy3}), the bulk action $S$ is invariant upto surface
terms:
\begin{equation}
\delta S=\frac{i}{\gamma}\int_{\partial\cal M}2F^{+-}\eta_{-}
\lambda_+ + F^{\perp-}\eta_{-}\lambda_{-}\, .
\label{varisusy}
\end{equation}
The first term vanishes because $F^{+-}=0$ on $\partial{\cal M}$
and the second term vanishes because $\lambda_{-}=0$ on
$\partial{\cal M}$. Including $\delta A_\mu=\bar\eta\partial_\mu\psi$
in the {\sc susy} transformations, 
we obtain an additional surface variation:
\begin{equation}
\delta S=-\frac{ik}{4\pi}\int_{\partial\cal M}F^{+-}\eta_{-}\psi_{+}
\end{equation}
which again vanishes because $F^{+-}=0$ on $\partial{\cal M}$.  
Hence we conclude that for {\sc n} boundary conditions there are
no dynamical edge states, neither bosonic nor fermionic, and 
{\sc susy} is maintained in the bulk.
\begin{itemize}
\item{\bf C boundary conditions:}
\end{itemize}
In Section\,2 we found that bosonic edge states are induced
if we fix $\bar A_+$ and $\bar\Pi^{\perp-}$ on $\partial{\cal M}$ and
include the surface term $S_2$ in the action. Since $\bar A_{-}$ is 
allowed to vary, rather than fixing $\lambda_{-}=0$ in (\ref{rsusy2}), 
we should instead fix $\lambda_+ =0$ on $\partial{\cal M}$
in order to extremize $S_3$. Fixing $\lambda_+$ on
$\partial{\cal M}$ in turn implies by (\ref{rsusy3}) that
$F_{\perp+}=0$ on $\partial{\cal M}$. Under the restricted
{\sc susy} transformations (\ref{rsusy1})
-- (\ref{rsusy3}), the bulk action $S$ is again invariant upto 
the surface terms given in (\ref{varisusy}). These terms now vanish, 
however, by virtue of the {\sc c} boundary conditions. The first term
vanishes because $\lambda_+=0$ on $\partial{\cal M}$. The second term
vanishes because $F^{\perp-}=F_{\perp+}=0$ on $\partial{\cal M}$.
Including $\delta A_\mu=\bar\eta\partial_\mu\psi$ in the {\sc susy}
transformations, we obtain an additional surface variation:
\begin{equation}
\delta S = -\frac{ik}{4\pi}\int_{\partial\cal M}\partial_+\theta
\eta_{-}\partial_{-}\psi_{+}
\end{equation}
which does not vanish. To maintain {\sc susy} we must therefore
supplement the bosonic action (\ref{bosonic}) with the surface term
\begin{equation}
S_{\tiny F}=\frac{k}{8\pi}\int_{\partial\cal M}\psi_{+}\partial_{-}
\psi_{+}
\ \ \ \ \mbox{with}\ \ \ \ \delta\psi_{+}\Big|_{\partial\cal M}
=i\partial_{+}\theta\eta_{-}\, ,
\label{sfermion}
\end{equation}
which may be recognized as the chiral fermionic action of superstring
theory.
\begin{itemize}
\item[ii)] $\eta_{-}=0:$
\end{itemize}
In this case the restricted {\sc susy} transformations are:
\begin{eqnarray}
\delta\bar A_+ &=& \!\!-i\eta_+\lambda_+ \label{rsusy4}\\
\delta\bar A_- &=&  i\eta_-\lambda_- =0    \label{rsusy5}\\
\mbox{$\delta\lambda\,\,\,$}  &=& -2F_{+-}\pmatrix{0 \cr \,\,\eta_+} - F_{\perp-}
\pmatrix{\,\eta_{+} \cr 0}\, . \label{rsusy6}
\end{eqnarray}
Fermionic edge states of opposite chirality are obtained if we impose
$\tilde{\mbox{\sc c}}$ boundary conditions.
\begin{itemize} 
\item {\bf $\tilde{\mbox{C}}$ boundary conditions:}
\end{itemize}
Recall that for $\tilde{\mbox{\sc c}}$ boundary conditions we fix
$\bar A_{-}$ and $\bar\Pi^{\perp+}$ on $\partial{\cal M}$ and add
$\tilde S_2$  to the action. Since $\bar A_+$ is allowed to vary in
(\ref{rsusy4}), rather than fixing $\lambda_{+}=0$ on $\partial
{\cal M}$, we should instead fix $\lambda_{-}=0$ on $\partial{\cal M}$
in order to extremize $S_3$. Fixing $\lambda_{-}$ on $\partial
{\cal M}$ in turn implies by (\ref{rsusy6}) that $F_{\perp-}=0$ on 
$\partial{\cal M}$. Under the restricted {\sc susy} transformations
(\ref{rsusy4}) -- (\ref{rsusy6}), the bulk action is invariant upto
surface terms:
\begin{equation}
\delta S = -\frac{i}{\gamma}\int_{\partial\cal M}2F_{+-}\eta_+
\lambda_{-} + F^{\perp+}\eta_+\lambda_+\, .
\end{equation}
The first term vanishes because $\lambda_{-}=0$ on $\partial{\cal M}$
and the second term vanishes because $F^{\perp+}=F_{\perp-}=0$ on
$\partial{\cal M}$. Under $\delta A_\mu=\bar\eta\partial_\mu\psi$
we obtain a non-vanishing surface term:
\begin{equation}
\delta S=-\frac{ik}{4\pi}\int_{\partial\cal M}\partial_{-}\theta
\eta_+\partial_+\psi_{-}\, .
\end{equation}
To maintain {\sc susy} we must add the fermionic action
\begin{equation}
\tilde{S}_{\tiny F}=\frac{k}{8\pi}\int_{\partial\cal M}\psi_{-}
\partial_+\psi_{-}
\ \ \ \ \mbox{with}\ \ \ \ \delta\psi_{-}\Big|_{\partial\cal M}
=i\partial_{-}\theta\eta_{+}\,,
\end{equation}
whose chirality is opposite to the action (\ref{sfermion}) obtained 
using {\sc c} boundary conditions. This agrees with our earlier
observation that parity interchanges {\sc c} and 
$\tilde{\mbox{\sc c}}$ boundary conditions.

\newsection{String constructions} 

Recall that the Neveu-Schwarz--Ramond model \cite{nsr}, 
which exhibits 2D worldsheet
{\sc susy}, contains both left and right movers on the string
worldsheet. We therefore consider a cylindrical topological membrane
with left ({\sc l}) and right ({\sc r}) boundaries. Depending on the 
boundary conditions, the bulk {\sc tmgt} may then induce left movers
on {\sc l} and right movers on {\sc r}.
The full left-right symmetric string worldsheet is obtained by 
gluing the separate left and right worldsheets traced out by {\sc l}
and {\sc r}. In this letter we shall only consider closed oriented 
strings (i.e. type II and heterotic) and so {\sc l} and {\sc r} must 
have the same orientation upon gluing. To obtain both left and right movers 
we must impose {\sc c} boundary conditions on both {\sc l} and 
{\sc r}. We denote this construction by {\sc cc}. Note that the string
construction based on {\sc c}$\tilde{\mbox{\sc c}}$ boundary
conditions has no (global) {\sc susy} since we must simultaneously
set $\eta_{-}=\eta_+=0$.

Type I theories, on the other hand, are based on unoriented open and
closed strings. We should obtain unoriented closed strings in the 
{\sc tm} approach by gluing {\sc l} and {\sc r} with opposite
orientations. We hope to give the precise details of this gluing 
in a future publication. We should also point out that an open string
construction has been obtained from pure Chern-Simons gauge theory
on a three dimensional ${\bb Z}_2$ orbifold \cite{horava1994a}. Motivated by
string duality, it would  be interesting to consider the full {\sc
tmgt} on a ${\bb Z}_2$ orbifold.

\begin{figure}[p]
\centerline{\psfig{figure=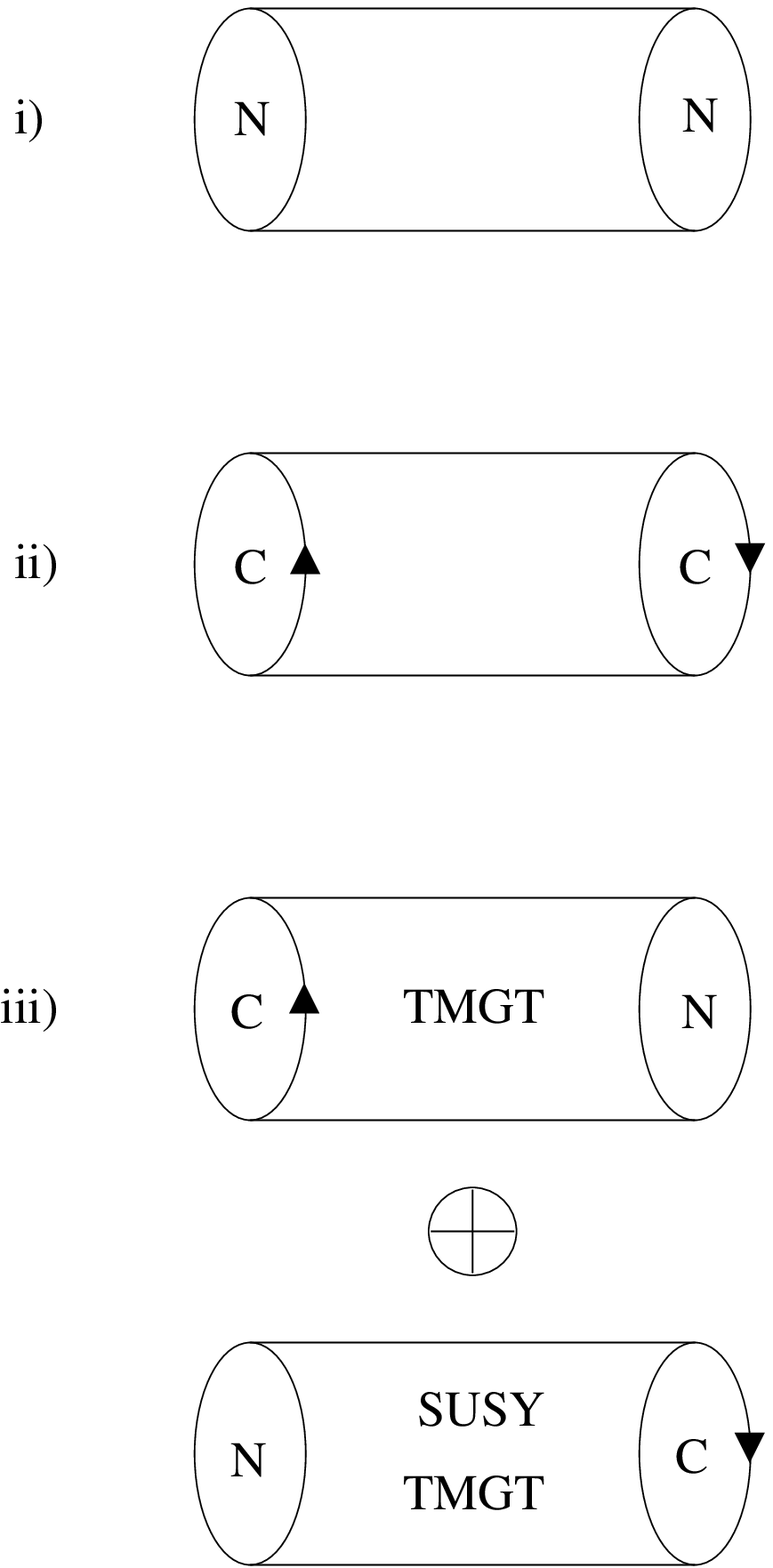,width=0.5\textwidth}}
\vspace{2cm}
\caption[]{String constructions.\ i) no-string \ ii) oriented closed 
string  \ iii) heterotic string.}
\label{strings}
\end{figure}

The crucial idea which permits a heterotic construction is that we can 
independently excite the left and right sectors by choosing different
boundary conditions on the left and right boundaries. Recall that
the right-moving sector of the heterotic string is the
10-dimensional superstring, while the left-moving sector is the
26-dimensional bosonic string which has been compactified to
ten dimensions. When the left and right sectors are put
together, they produce a self-consistent, ghost-free,
anomaly-free, one-loop finite theory. The seemingly hybrid
nature of this construction becomes more natural in the
{\sc tm} approach. Suppose the topological membrane has a
semi-simple gauge group $G_{\tiny L}\times G_{\tiny R}$ 
where $G_{\tiny L}$ represents ordinary {\sc tmgt} and
$G_{\tiny R}$ represents {\sc susy tmgt}. The heterotic 
construction is obtained if we then impose {\sc cn} 
boundary conditions on $G_{\tiny L}$ and {\sc nc} boundary conditions 
on $G_{\tiny R}$. It is appealling that the different left 
and right sectors of the heterotic string have a common
geometric origin in the context of the topological
membrane. Each of the above string constructions are 
illustrated in Fig.\,\ref{strings}.

\newsection{Bulk spectrum and Casimir energy} 

We now calculate the spectrum of the topological membrane
in the bulk for each of the above string constructions. 
We consider the cylindrical membrane $I\times S^1$, where
the interval $I$ has length $L$ and the two circular boundaries have
radii $R$. For simplicity, we shall only consider the case where 
$R >> L$ which, in the limit  $R \rightarrow \infty$, is equivalent to
two infinitely long parallel wires held a distance $L$ apart.
The coordinate system is chosen such that the two wires are
parallel to the $x_1$ spatial axis and lie, respectively,
on $x_2=0$ and $x_2=L$. Eq.\,(\ref{wave2}) shows that the gauge field
in the bulk has a plane-wave spectrum. Likewise, (\ref{varif})
gives the fermionic equations of motion $(\not\!\!\partial-m)
\lambda=0$ which, multiplied by $(\not\!\!\partial+m)$, shows that
$\lambda_{+}$ and $\lambda_{-}$ each satisfy the wave equation
$(\partial^2-m^2)\lambda_\pm=0.$ 
\begin{itemize}
\item {\bf NN (no-string) boundary conditions:}
\end{itemize}
We impose the gauge condition $A_+=0$
so that on both boundaries the value of $A_+$ is fixed.
Plane-wave solutions to (\ref{wave2}) can then be constructed from the
gauge potential ($A_+=0$):
\begin{equation}
A_i(x)=a_i(x_2)e^{i(\omega t-k_1x_1)}\ \ \ \ \ \ \ i=-,2
\end{equation}
where the explicit form of $a_i(x_2)$ is determined by the boundary
conditions. {\sc nn} boundary conditions are that $A_{-}=0$ at both
$x_2=0$ and $x_2=L$, which are satisfied by
\begin{equation}
a_-(x_2)=\sin k_2x_2 \ \ \ \ \ \mbox{with} \ \ \ \ \ k_2=n\pi/L\, .
\label{a2}
\end{equation}
In order to have consistent commutation relations, we must also impose
$\Pi^{2-}\neq 0$ on $\partial{\cal M}$. Substituting Eq.\,(\ref{a2}) into 
(\ref{mom}) gives 
\begin{eqnarray}
\Pi^{2-} &=& -\frac{1}{\gamma}F^{2-}+\frac{k}{8\pi}\epsilon^{2-+}A_+
\nonumber\\
 &=& \frac{i}{\gamma}\big(\omega-k_1\big)\,a_2(x_2) e^{i(\omega t-k_1x_1)}\, ,
\label{momexpl}
\end{eqnarray}
which is indeed non-vanishing at $x_2=0$ and $L$ if 
$a_2(x_2)=\cos k_2x_2$. Thus the gauge potential satisfying
{\sc nn} boundary conditions is
\begin{equation}
A_i(x)=\pmatrix{\sin k_2x_2\cr\cos k_2x_2}e^{i(\omega t -k_1x_1)}
\ \ \ \ \ \mbox{with} \ \ \ \ \ k_2=n\pi/L\, .
\label{Ann}
\end{equation}
The corresponding dual field strength is
\begin{equation}
F^\mu=\epsilon^{\mu\alpha\beta}\partial_\alpha A_\beta=
\pmatrix{(k_2-ik_1)\cos k_2x_2\cr (k_2-i\omega)\cos k_2x_2\cr 
 (ik_1-i\omega)\sin k_2x_2}e^{i(\omega t-k_1x_1)}
\label{fnn}
\end{equation}
with spectrum
\begin{equation}
\omega_n^2=\left(\frac{n\pi}{L}\right)^2 + k_1^2 + m^2 
\label{nnspec}
\end{equation}
 Note that {\sc nn} boundary conditions correspond to the perfectly
conducting case (see discussion after (\ref{perfectconductor})) 
since $F^2=0$ on $\partial\cal{M}$. 
For the bulk fermions, {\sc nn} boundary conditions are $\lambda_{-}
=0$ (but $\lambda_+\neq 0$) at both $x_2=0$ and $x_2=L$. Hence
\begin{equation}
\pmatrix{\lambda_{-} \cr \lambda_+} = \pmatrix{\sin k_2x_2 \cr
\cos k_2x_2}e^{i(\omega t-k_1x_1)}
\ \ \ \ \ \mbox{with} \ \ \ \ \ k_2=n\pi/L\, ,
\end{equation}
and so the fermionic spectrum indeed matches the bosonic spectrum
(\ref{nnspec}).
\begin{itemize}
\item{\bf CC (oriented closed superstring) boundary conditions:}
\end{itemize}
{\sc susy} edge states are induced on both boundaries if we
impose {\sc cc} boundary conditions, namely $A_{-}\neq 0$ at both
$x_2=0$ and
$x_2=L$. This occurs if
\begin{equation}
a_{-}(x_2)=\cos k_2x_2  \ \ \ \ \ \mbox{with} \ \ \ \ \ k_2=n\pi/L\, .
\end{equation}
Simultaneously, we must fix $\Pi^{2-}=0$ on both boundaries. Hence we
must have $a_2(x_2)=\sin k_2x_2$ in (\ref{momexpl}).
The gauge potential satisfying {\sc cc} boundary conditions is
\begin{equation}
A_i(x)=\pmatrix{\cos k_2x_2\cr\sin k_2x_2}e^{i(\omega t -k_1x_1)}
\ \ \ \ \ \mbox{with} \ \ \ \ \ k_2=n\pi/L\, .
\label{Acc}
\end{equation}
The corresponding dual field strength is
\begin{equation}
F^\mu= -\pmatrix{(ik_1+k_2)\sin k_2x_2\cr
      (i\omega+k_2)\sin k_2x_2\cr (i\omega-ik_1)\cos k_2x_2}e^{i(\omega 
       t-k_1x_1)}
\label{fcc}
\end{equation}
with spectrum indentical to Eq.\,(\ref{nnspec}). {\sc cc} boundary 
conditions on the fermions are $\lambda_{+}=0$ (but $\lambda_-\neq 0$)
at $x_2=0$ and $x_2=L$. Hence
\begin{equation}
\pmatrix{\lambda_{-} \cr \lambda_+} = \pmatrix{\cos k_2x_2 \cr
\sin k_2x_2}e^{i(\omega t-k_1x_1)}
\ \ \ \ \ \mbox{with} \ \ \ \ \ k_2=n\pi/L\, ,
\end{equation}
which verifies that the {\sc cc} spectrum is indeed supersymmetric.
\begin{itemize}
\item{\bf NC (heterotic string) boundary conditions:}
\end{itemize}
The gauge potential satisfying {\sc nc} boundary conditions is 
given by (\ref{Ann}) with $k_2=(n+1/2)\pi/L$. As before,
there will be no bosonic edge states induced on the boundary $x_2=0$ since
$A_{-}(x_2=0)=0$. However, because now the spectrum is half integer,
$A_{-}(x_2=L)\neq 0$ and so edge states will be induced on the boundary
$x_2=L$. Moreover, we have $\Pi_{2-}(x_2=L)=0$, which ensures
the action (\ref{action}) has a well-defined classical limit.
The bosonic spectrum for the {\sc nc} case is
\begin{equation}
\omega_n^2=\left(\left(n+\mbox{$\frac{1}{2}$}\right)\frac{\pi}{L}\right)^2
  +k_1^2+m^2 \ \ \ \ \ \ \ \ (\mbox{\sc{nc \& cn}})\,.
\label{ncspec}
\end{equation}
The {\sc nc} boundary conditions for fermions are mixed. We must
set $\lambda_{-}=0$ (but $\lambda_{+}\neq 0$) at $x_2=0$ and
$\lambda_{+}=0$ (but $\lambda_{-}\neq 0$) at $x_2=L$.
Hence
\begin{equation}
\pmatrix{\lambda_{-} \cr \lambda_+} = \pmatrix{\sin k_2x_2 \cr
\cos k_2x_2}e^{i(\omega t-k_1x_1)}
\ \ \ \ \ \mbox{with} \ \ \ \ \ k_2=\left(n+\mbox{$\frac{1}{2}$}
\right)\mbox{$\frac{\pi}{L}$}\, ,
\end{equation}
which matches the bosonic spectrum (\ref{ncspec}).

We now calculate the Casimir energy for each of the above string
constructions. Since the {\sc nn} and {\sc cc} spectra are 
supersymmetric, we immediately know that the no-string ({\sc nn})
and normal closed string ({\sc cc}) constructions have zero Casimir
energy. Likewise, the supersymmetric right sector of the heterotic 
({\sc nc}) construction has zero Casimir energy. However, the bosonic
({\sc cn}) left sector is not supersymmetric and so may have a non-zero
Casimir energy. Defined as the infinite sum over zero modes, this
Casimir energy is
\begin{equation}
E_0=\frac{1}{2L}\sum_{n=-\infty}^\infty\int\frac{dk_2}{2\pi}\left\{\Big(
(n+\mbox{$\frac{1}{2}$})\mbox{$\frac{\pi}{L}$}\Big)^2+k_2^2+m^2
\right\}^{1/2}
\label{einf}
\end{equation}
Replacing the power $1/2$ in
the integrand with $-s/2$, where $s$ is a complex variable,
the integral can be evaluated as the analytic continuation
of the Euler beta function, yielding
\begin{equation}
E_0=\lim_{s\rightarrow -1}\,\frac{\pi^{\frac{1-2s}{2}}}{4L^{2-s}}\,
    \frac{\Gamma(\frac{s-1}{2})}{\Gamma(\frac{s}{2})}
    \sum_{n=-\infty}^\infty\left\{(n+1/2)^2+q\right\}
    ^{\frac{1-s}{2}}\, ,
\end{equation}
where $q=(mL/\pi)^2$. The analytic continuation of this series, 
known as an  Epstein-Hurwitz zeta function, is \cite{elizalde1994a}:
\begin{eqnarray}
\sum_{n=-\infty}^\infty\left\{(n+\delta)^2+q\right\}^{-t}
 &=& \pi^{1/2}\frac{\Gamma(t-1/2)}{\Gamma(t)}q^{(1/2-t)}\,+\,
     \frac{4\pi^t}{\Gamma(t)}q^{1/4-t/2}\nonumber\\
 & & \times\sum_{n=1}^\infty n^{t-1/2}\cos(2\pi n\delta)K_{t-1/2}
     (2\pi n\sqrt{q})
\end{eqnarray}
where $K_\nu$ is the modified Bessel function of the second
kind. Substituting $t=-(1-s)/2$ and then taking the limit 
$s\rightarrow -1$, we obtain the regularized Casimir energy density
\begin{equation}
E_0=-\frac{1}{6\pi}\frac{1}{L^3}\left\{(mL)^3+3\pi^{-1/2}(mL)^{3/2}
    \sum_{n=1}^\infty\frac{(-1)^n}{n^{3/2}}\,K_{-3/2}(2nmL)\right\}\,.
\label{energy}
\end{equation}
The first term in (\ref{energy}) is independent of the separation $L$ 
between the wires and can be dropped. Rewriting the second
term using the identity \cite{erdelyiET}
\begin{equation}
K_{-3/2}(z)=K_{3/2}(z)=\sqrt{\mbox{$\frac{\pi}{2}$}}e^{-z}z^{-3/2}(1+z)
\end{equation}
gives a positive Casimir energy
\begin{equation}
E_0=-\frac{1}{8\pi}\frac{1}{L^3}\sum_{n=1}^\infty\frac{(-1)^n}{n^3}
e^{-2nmL}(1+2nmL)\, .
\label{ereg}
\end{equation}
The resulting Casimir force (per unit length) is
\begin{eqnarray}
F(L)&=&(-\partial/\partial L)\,\{L\times E_0(L)\} \nonumber \\
    &=&-\frac{1}{2\pi}\frac{1}{L^3}\sum_{n=1}^\infty(-1)^ne^{-2nmL}
       \left\{\frac{1}{2n^3}+\frac{mL}{n^2}+\frac{(mL)^2}{n}\right\}\,.
\label{force}
\end{eqnarray}
We see that the  heterotic construction is dynamically distinguished
by the fact that it has a repulsive Casimir force and so, in flat
space, the membrane expands and will continue to do so until we can no
longer assume $R>>L$. Whether or not the effects of the cylindrical 
geometry will stabilize this expansion depends on the 
geometrodynamics of the topological membrane \cite{carlip1991a,grav}. 
 We note that  modular invariance on the worldsheet should also be
 connected with the gravitational sector of the topological membrane 
 and the cancellation of the global gravitational anomaly  in the
bulk. We hope to address these issues in future work.

\newsection{Conclusions}

 In this letter we have shown how different boundary conditions 
in the topological membane theory lead to different string
constructions. For certain boundary conditions, it was shown how
some bosonic gauge degrees of freedom become dynamical on the membrane
boundary, which was given an interpretation as the string worldsheet. 
Worldsheet fermions were incorporated in this picture by using a 
supersymmetric topological membrane. The appeal of this approach is
that the different left and right sectors of the heterotic string
have a clear geometrical interpretation as a specific choice of
membrane boundary conditions --- and this choice is only possible 
in the full {\sc tmgt}, and not in a pure Chern-Simons theory. 
By calculating the spectrum in the bulk, we then showed that the
supersymmetric oriented closed string ({\sc cc}) and no-string 
({\sc nn}) constructions have zero Casimir energy. The heterotic 
({\sc nc}) construction, on the other hand, has a positive Casimir 
energy. Although the physical implications of this result are not yet
clear to us, it is apparent, at least from the bulk perspective,
that the {\sc tm} approach distinguishes the heterotic construction.
In turn, this could have interesting consequences for string duality 
and the proposal that all ten-dimensional string theories have
a common eleven-dimensional origin, i.e. $M$-theory. 

Ho\v rava and 
Witten \cite{horavaET} have argued that the strong coupling limit
of the ten-dimensional $E_8\times E_8$ heterotic string is 
$M$-theory compactified on ${\bb R}^{10}\times S^1/{\bb Z}_2$.
Their construction shares some amusing similarities with the
{\sc tm} approach, although we emphasize that the two approaches 
are fundamentally 
different. In the Ho\v rava-Witten approach, space-time ideas such
as eleven-dimensional supergravity play a prominent role in
determining the structure of the theory. For instance,
the $E_8\times E_8$ heterotic ``string'' is seen to be a cylindrical 
two-brane with one boundary attached to each boundary of space-time. 
Cancellation of gauge and gravitational anomalies then requires there 
to be one $E_8$ gauge group on each boundary of spacetime. In
contrast, the emphasis of the {\sc tm} approach is on the worlsheet
properties of the open topological membrane. Nevertheless, it would
be interesting to see if the extra  dimension of {\sc tm} theory 
could somehow be embedded in eleven-dimensional $M$-theory.

It has also very recently been shown that Chern-Simons couplings arise
in the effective world-volume action for a type IIA superstring
Dirichlet two-brane \cite{csbrane}, which is supposed to descend 
from the membrane of $M$-theory. Moreover, the five-brane of
$M$-theory has been interpreted  as a $D$-brane of an open 
supermembrane in eleven dimensions \cite{dbrane}. Motivated
by these results, it would be interesting to see if the
topological membrane discussed in this letter has a role in 
$M$-theory.

\section*{Acknowledgements}

We would like to thank Shaun Cooper, Kai-Ming Lee,  Graham Ross
and Paul Townsend for useful discussions. One of us 
{\sc l.c.} acknowledges financial support from the University of 
Canterbury, New Zealand.

\newpage

\end{document}